# VERIFIABLE DATA SHARING SCHEME FOR DYNAMIC MULTI-OWNER SETTING

Jing Zhao and Qianqian Su

Department of Computer Science and Technology, Qingdao University,
Qingdao, China

## ABSTRACT

*One of scenarios in data-sharing applications is that files are managed by multiple owners, and the list of file owners may change dynamically. However, most existing solutions to this problem rely on trusted third parties and have complicated signature permission processes, resulting in additional overhead. Therefore, we propose a verifiable data-sharing scheme (VDS-DM) that can support dynamic multi-owner scenarios. We introduce a management entity that combines linear secret-sharing technology, multi-owner signature generation, and an aggregation technique to allow multi-owner file sharing. Without the help of trusted third parties, VDS-DM can update file signatures for dynamically changing file owners, which helps save communication overhead. Moreover, users independently verify the integrity of files without resorting to a third party. We analyze the security of VDS-DM through a security game. Finally, we conduct enough simulation experiments and the outcomes of experimental demonstrate the feasibility of VDS-DM.*

## KEYWORDS

*Security, Data Sharing, Dynamic Multi-Owner, Verification*

## 1. INTRODUCTION

Thanks to the fast growth of cloud computing [1, 2, 3, 4], companies and individuals are able to store files on cloud servers for easy sharing. Although cloud computing offers many conveniences, it also brings several security risks [5, 6]. First, there is a danger of file privacy leakage since files may contain sensitive information and cloud servers cannot be completely trusted. Second, when file is kept in the cloud, the file owner loses physical control over the file, increasing the risk of illegal access. Naturally, file confidential can be achieved via traditional symmetric and asymmetric encryption techniques. However, these methods enable one-to-one access control rather than flexible and controlled authorized access. In addition, when there are many files, these methods suffer from drawbacks such as multiple copies of ciphertext, high encryption overhead and complicated key management [7, 8]. Fortunately, a potential solution to the above problems has emerged with the emergence of the ciphertext-policy attribute-based encryption (CP-ABE) scheme [9]. In CP-ABE, the file owner determines the set of authorized users by establishing an attribute-based access policy. Only users whose attributes satisfy the access policy can achieve decryption of the ciphertext. To achieve multi-user-oriented sharing in CP-ABE, the file owner just has to encrypt the file once. Therefore, CP-ABE can leverage attributes to achieve one-to-many access control by performing only one time encrypt operation [10, 11, 12, 13].

Most of CP-ABE based file sharing schemes are designed for single-file-owner setting, where the files are owned and managed by a single user [14, 15, 16, 17]. However, a multi-owner setting, in which files are managed by many owners, is equally typical. In contrast to single-file-owner, sharing files under a multi-owner setting require everyone's signature for permission. Obviously, schemes designed for single-file-owner setting are not suitable for multi-owner setting. This is due to the fact that the latter has to address not just the issue of dynamic user changes, but also whether permissions are obtained from all file owners. One option to address the above issues is to appoint a multi-owner representative as manager with the responsibility of defining file access policies and ensuring file confidentiality. Different from the single-file-owner scenario, the multi-owner scenario will face new problems. First, the manager must obtain each owner's approval before the file shared. After that, the ciphertext and permissions can be uploaded to the cloud. To reduce storage and communication overhead, the manager has to aggregate multiple permissions into one. Second, the file owners may leave or join, both of which have an impact on the permission of the file. Therefore, the manager needs to perform file updates with the least amount of overhead as possible. Third, during the access phase, the user expects the file supplied by the cloud server to be correct. However, the cloud server could give the user incomplete or incorrect file due to factors like software, hardware, or interest, [18, 19, 20, 21]. Therefore, the user needs to have the ability to check the integrity of the results.

Due to the prevalence of multi-owner setups, researchers are increasingly interested in how to design solutions for multi-owner scenarios. After in-depth research, we found that although related work has been proposed, there are still some problems that have not been well solved well. First, when the membership of file owner's changes (such as join or leave), most existing solutions require the manager to complete file updates with the assistance of a trusted third party. To enable the renewal of multi-user signatures, the third party needs to redistribute the public and private keys for the manager. This method increases additional communication overhead [20]. Second, most of the existing schemes utilize the integrity proof method for file integrity verification, which is complex and requires the help of a third party [20, 21].

### 1.1. Our Contribution

Motivated by the above issues, we design a verifiable data sharing scheme based on CP-ABE with dynamically multi-owner setting (shorted as VDS-DM). We introduce a management entity that combines linear secret-sharing technology, multi-owner signature generation, and an aggregation technique to allow multi-owner file sharing. Without the help of trusted third parties, VDS-DM can update file signatures for dynamically changing file owners, which helps save communication overhead. For the verifiability of the shared file, users can independently verify the integrity of the shared file without resorting to a trusted third party. Additionally, we analyze the security of VDS-DM through a formal security game. Finally, we carry out enough simulation experiments and the outcomes of experimental demonstrate the feasibility of VDS-DM.

## 2. RELATED WORK

Waters and Sahai [22] proposed the first attribute-based encryption (ABE) scheme, which is considered as an extension of identity-based encryption (IBE) [23]. In ABE schemes, a set of attributes is used as the user's identity. Only users whose attributes satisfy the specified access policy can obtain the plaintext of the encrypted file. This property of ABE has led to its increasing interest in data sharing applications. Later, Goyal et al. [24] proposed the first key-policy ABE (KP-ABE) scheme and Bethencourt et al. [9] proposed the first ciphertext-policy

ABE (CP-ABE) scheme. Due to the fact that CP-ABE supports data owners to set access policies, most of the subsequent studies on data sharing schemes are conducted based on CP-ABE.

In the next section, we mainly consider the data sharing schemes in the multi-owner scenario. In CP-ABE schemes that support multi-owner setting, Miao et al. [25] designed a verifiable keyword search scheme for encrypted data using multiple signature techniques. Miao et al. [26] proposed a CP-ABKS scheme based on privacy protection and implemented the traceability function of malicious users. Moreover, Zhang et al. [21] achieved multi-keyword search and verifiability of search results with guaranteed efficiency. However, the above schemes are implemented in static multi-owner setting, without considering dynamic multi-owner setting. In other words, the above schemes cannot be used directly to solve the problem if the file owners are added or deleted. Recently, Miao et al. [20] proposed a verifiable fine-grained keyword search scheme that supports dynamic multi-owner setting. However, their scheme requires interaction with a trusted third-party entity when performing update operation, which increases the time overhead.

Another issue in multi-owner data sharing is that the cloud service is in practice an incomplete trusted entity. Cloud servers may return incomplete or incorrect search results to users due to interest issues or software and hardware failures. Sun et al. [27] implemented the search result verification function to some extent by using Bloon filters. Due to the high false positive rate of the Bloon filter, the search results will not be verified accurately. Miao et al. [20] and Zhang et al. [21] used signatures of files to achieve verification of the integrity of search results, but the process of forming signatures is complicated and requires interaction with the cloud service at the time of verification.

## 3. PRELIMINARIES

### 3.1. Bilinear pairing

Let $\mathbb{G}, \mathbb{G}_\mathbb{T}$ be two multiplicative cyclic groups of prime order $p$, and $g$ is the generator of $\mathbb{G}$. The bilinear mapping function $e: \mathbb{G} \times \mathbb{G} \to \mathbb{G}_\mathbb{T}$ has the following three properties:

1) Bilinearity: $e(\mu_1^{\varphi_1}, \mu_2^{\varphi_2}) = e(\mu_1, \mu_2)^{\varphi_1 \varphi_2}$ for all $\mu_1, \mu_2 \in \mathbb{G}, \varphi_1, \varphi_2 \in \mathbb{Z}_p$.
2) Non-degeneracy: For all $\mu_1, \mu_2 \in \mathbb{G}, e(\mu_1, \mu_2) \neq 1$.
3) Computability: For all $\mu_1, \mu_2 \in \mathbb{G}, e(\mu_1, \mu_2)$ can be computed efficiently.

### 3.2. Access structure

$A = \{A_1, A_2, \cdots, A_n\}$ is the set of $n$ attributes. $\mathbb{A}$ represents an access structure. For any $B, C$, if $B \subseteq A$ and $B \subseteq C$, then $C \in \mathbb{A}$, then the collection $\mathbb{A} \subseteq 2^A$ is monotone. We can say that the collection $\mathbb{A}$ of non-empty subsets of $A$ is a monotone access structure. The set in $\mathbb{A}$ is called an authorized set and the opposite is called an unauthorized set.

### 3.3. Decisional parallel bilinear Diffie-Hellman exponent assumption

Let $\mathbb{G}$ be a group with order $p$ and $g$ is the generator of $\mathbb{G}$. Choose $a, s, b_1, \cdots, b_q \in \mathbb{Z}_p$. Given a tuple:

$$\vec{y} = \left(g, g^s, g^a, \cdots, g^{a^q}, g^{a^{q+2}}, \cdots, g^{a^{2q}}, \forall_{1 \leq j \leq q} g^{sb_j}, \right.$$
$$\left. g^{a/b_j}, \cdots, g^{a^q/b_j}, g^{a^{q+2}/b_j}, \cdots, g^{a^{2q}/b_j}, \forall_{1 \leq k,j \leq q, k \neq j} g^{asb_k/b_j}, \cdots, g^{a^q sb_k/b_j}\right).$$

There does not exist any probabilistic polynomial-time algorithm $\mathcal{B}$ that distinguishes with non-negligible probability between $e(g,g)^{a^{q+1}s} \in \mathbb{G}_T$ and a random element $R \in \mathbb{G}_T$. The algorithm $\mathcal{B}$, which returns the result $z \in \{0,1\}$, has advantage $\epsilon$ in solving the q-parallel BDHE problem in $\mathbb{G}$ if $\left| \Pr[\mathcal{B}(\vec{y}, e(g,g)^{a^{q+1}s}) = 0] - \Pr[\mathcal{B}(\vec{y}, R) = 0] \right| \geq \epsilon$.

### 3.4. Linear secret sharing scheme (LSSS)

Let $(\mathbb{M}, \rho)$ indicates an access policy $\mathbb{A}$, where $\mathbb{M}$ is a shared generator matrix with $l$ rows and $n$ columns. The function $\rho(i)(i \in [1, l])$ maps the $i$-th row in $\mathbb{M}$ to an attribute in $\mathbb{A}$. The linear secret sharing scheme consists of two parts: share generation and secret reconstruction.

1) Share generation. Secret $s \in \mathbb{Z}_p^*$, choose $y_2, \cdots, y_n \in \mathbb{Z}_p^*$ at random. Set a vector $\vec{v} = (s, y_2, \cdots, y_q)$ to compute $\lambda_i = \mathbb{M}_i \cdot \vec{v}$. $\lambda_i$ is the valid shared value of the secret $s$ corresponding to $\rho(i)$.
2) Secret reconstruction. Any authorized set $S \in \mathbb{A}, I \subset \{1, 2, \cdots, l\}$ is defined as $I = \{i : \rho(i) \in S\}$. There exists a set of constants $\{\omega_i \in \mathbb{Z}_p\}_{i \in I}$ that can be found in polynomial time. By computing $\sum_{i \in I} \omega_i \lambda_i = s$ can recover the secret $s$.

## 4. PROBLEM FORMULATION

In this section, we present the system model, algorithm definition, security model and design goals of the proposed scheme.

### 4.1. System Model

As shown in Figure 1, the system model consists of five entities: Trusted Authorization (**TA**), Date Manager (**DM**), Data Owners (**DOs**), Cloud Service Provider (**CSP**), Data User (**DU**). The details of each entity are as follows:

- **TA** is a trusted authorization center. **TA** is responsible for initializing the system and generating the secret key for each **DU** (Step (1)).
- **DOs** is the owner of data. Each **DO** generate the signature to indicate approval of the file and sends the signature to **DM** (Step(2)).
- **DM** is a representative of all **DOs**. **DM** has own public/private key pair. **DM** first encrypts the file with the symmetric encryption method, then **DM** sets the access policy and uses to encrypt the symmetric key. After receiving the signatures of all **DOs**, **DM** aggregates them into a single signature. Finally, **DM** sends the file ciphertext, the key ciphertext, and the aggregated signature to **CSP** (Step (3)). When the number of **DOs** changes, **DM** issues an update key to **CSP** to update the aggregated signature (Step (8)).
- **CSP** is a semi-honest cloud server provider, which provides storage, search and update services. **CSP** stores the ciphertext sent by **DM**. When **DU** sends search request to it, **CSP** returns the search results to **DU** (Step (5)). Besides, **CSP** implements the signature update operation by using the update key sent by **DM**.
- **DU** is the user of data. **DU** first sends a file search request to **CSP** (Step (4)), then **DU** verifies the search results returned by **CSP** (Step (6)). If the verification fails, **DU** will no longer execute the decryption algorithm. Otherwise, the ciphertext can be decrypted only if the attributes of **DU** meet the access policy (Step (7)).

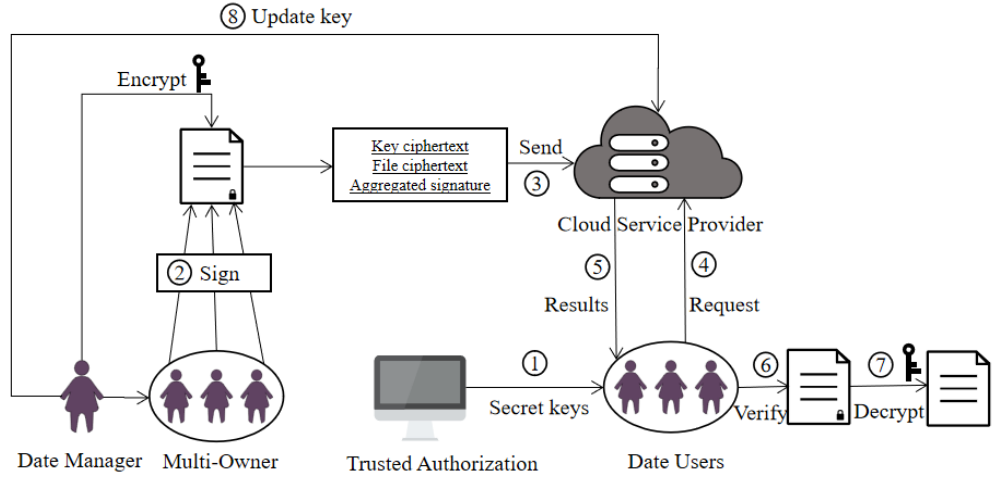

Figure 1. The system model of VDS-DM

## 4.2. Algorithm definition

Our proposed scheme includes the following algorithms:

- **Setup**$(1^\kappa)$. Given a secure parameter $\kappa$, **TA** runs this algorithm to output the public key $PK$ and the master key $MSK$.
- **KeyGen**$_{DU}(PK, MSK, S)$. Given the public key $PK$, the master key $MSK$ and **DU**'s attribute set $S$, **TA** runs this algorithm to output **DU**'s secret key $SK_u$.
- **KeyGen**$_{DM}(PK, \mathcal{O})$. Given the public key $PK$ and **DO**'s set $\mathcal{O} = \{O_1, \cdots, O_d\}$, **DM** outputs own public/private key pair $(PK_m, SK_m)$ and parameter $y_t (t \in [1, d])$ for each **DO**.
- **Enc**$(PK, (\mathbb{M}, \rho), F)$. Given the public key $PK$, matrix access structure $\mathbb{M}$, the function $\rho$ maps each row of the $\mathbb{M}$ to a attribute and file $F$, **DM** outputs file ciphertext $C_F$ and key ciphertext $CT_1$.
- **Sign**$(C_F, y_t)$. Given file ciphertext $C_F$ and parameter $y_t$, **DO**$_t$ outputs signature $\sigma_t$.
- **Agg**$(\{\sigma_t\})$. Given all **DOs**'s signature set $\{\sigma_t\}$, **DM** outputs an aggregated signature $\sigma$ and uploads the ciphertext $CT = \{C_F, CT_1, \sigma\}$ to **CSP**.
- **Verify**$(PK, CT^*, PK_m)$. Given the public key $PK$, the ciphertext $CT^*$ sent from **CSP** and **DM**'s public key $PK_m$, **DU** validates the integrity of $CT^*$.
- **Dec**$(CT^*, SK_u)$. Given the ciphertext $CT^*$ and **DU**'s secret key $SK_u$, Only **DU** whose attributes satisfy the access policy can use the secret key $SK_u$ to decrypt successfully. Authorized **DU** outputs the file $F$.
- **Update**$(SK_m)$. Given **DM**'s private key $SK_m$, **DM** outputs the update key $UPK$. **DM** sends it to **CSP** to implement the signature update operation.

## 4.3. Security Model

We require that the proposed scheme is secure against selected plaintext attacks in a selective model. We describe the security model by designing a security game played by the adversary $\mathcal{A}$ and the challenger $\mathcal{C}$.

*Initialization*: $\mathcal{A}$ selects a challenge access policy $P^*$ and sends $P^*$ to $\mathcal{C}$.
*Setup*: $\mathcal{C}$ executes the **Setup** algorithm to obtain the public key $PK$ and the master secret key $MSK$. Then, it sends the public key $PK$ to $\mathcal{A}$.

***Phase 1***: $\mathcal{A}$ makes a key generation query to $\mathcal{C}$, $\mathcal{C}$ executes the **KeyGen$_{DU}$** algorithm to generate a key $SK_u$ and sends it to $\mathcal{A}$. It is noted that the attribute set $S$ of attributes of $\mathcal{A}$ does not satisfy $P^*$.

***Challenge***: $\mathcal{A}$ submits two messages of the same length $M_0, M_1$ to $\mathcal{C}$. Then, $\mathcal{C}$ chooses a random bit $b \in \{0,1\}$ and runs the **Enc** algorithm to generate the ciphertext $CT_b^*$ under the challenge access policy $P^*$. Finally, $\mathcal{C}$ sends $CT_b^*$ to $\mathcal{A}$.

***Guess***: $\mathcal{A}$ outputs a guess $b' \in \{0,1\}$.
The advantage of the adversary $\mathcal{A}$ to winning this game is defined as:
$Adv_{\mathcal{A}}^{CPA} = \left|\Pr[b' = b] - \frac{1}{2}\right|$.

**Definition 1.** If there does not exist any probabilistic polynomial-time adversary $\mathcal{A}$ that can break the above selectively safe game with non-negligible probability, then VDS-DM is selectively safe.

### 4.4. Design goals

**File integrity**. If **CSP** returns an incomplete result to **DU**, **DU** can detect the error with a non-negligible probability.
**File privacy**. If and only if the attributes of **DU** satisfy the access policy set by **DM**, **DU** can decrypt the ciphertext.
**Update Correctness**. If the number of **DOs** changes, the ciphertext and owner's permission information can be updated correctly.

## 5. THE PROPOSED VDS-DM

In this section, after presenting the overview flow of the VDS-DM scheme, we describe in detail the algorithm construction involved in VDS-DM. Finally, we prove the security of the VDS-DM scheme.

### 5.1. Overview

As shown in Figure 2, **TA** runs **Setup** algorithm to realize system initialization and runs **KeyGen$_{DU}$** algorithm to generate the secret key for each **DU**. Each **DO** execute **Sign** algorithm to generate a signature to achieve approval of the file. **DM** runs **KeyGen$_{DM}$** algorithm to generate public/private key pair and publishes public key. **DM** performs symmetric encryption on the file and sets the access policy. **DM** uses the access policy to encrypt the symmetric key under **Enc** algorithm. Then **DM** aggregates all signatures into one signature by executing **Agg** algorithm. Finally, **DM** uploads the file ciphertext, the key ciphertext, and the aggregated signature to **CSP**. **DU** sends a search request to **CSP**, and **CSP** returns results to **DU**. Then **DU** executes **Verify** algorithm to verify the integrity of the results. If the verification passes, only **DU** whose attributes match the access policy can use **Dec** algorithm to decrypt successfully. When the number of **DOs** changes, **DM** runs **Update** algorithm to generate an update key without the help of a third party and uploads it to **CSP**. **CSP** uses the update key to complete the signature update operation.

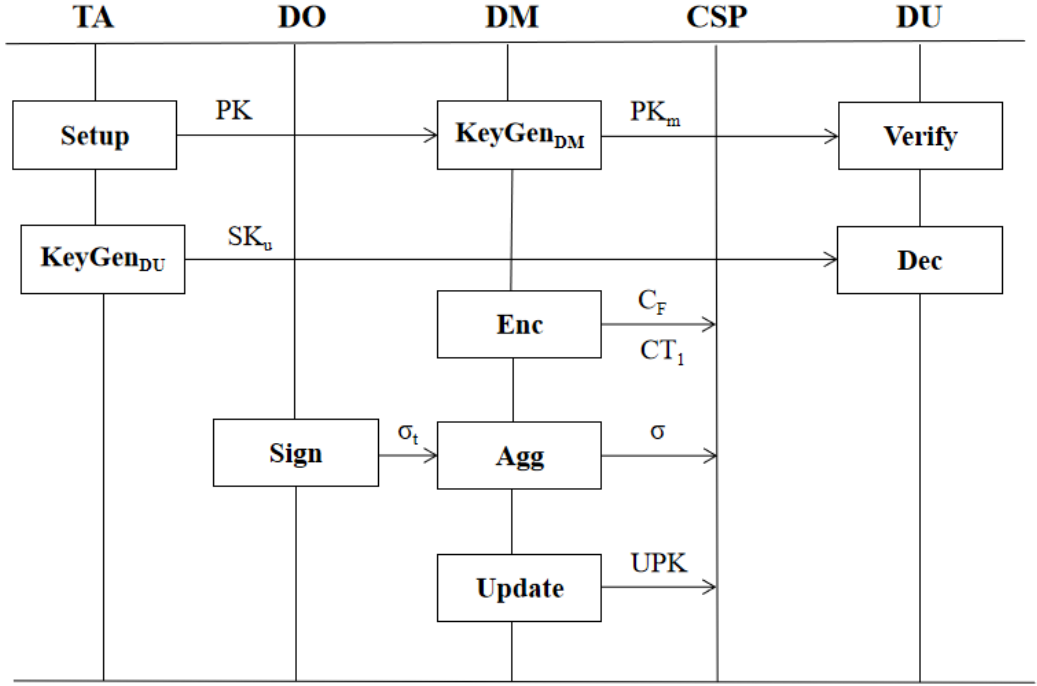

Figure 2. The system flow of VDS-DM

## 5.2. Construction of VDS-DM

The detailed algorithms of the VDS-DM are described as follows:

- **Setup**$(1^\kappa)$. Given a global public parameter $pp = (\mathbb{G}, \mathbb{G}_T, p, e, g)$, **TA** randomly selects elements $\alpha, \beta \in \mathbb{Z}_p$, and computes $e(g,g)^\alpha$ and $g^\beta$. **TA** defines a system attribute set $A$, and selects a random element $h_x \in \mathbb{G}$ for each attribute $x \in A$. **TA** chooses a anti-collision hash function $H_2: \{0,1\} \to \mathbb{G}$. **TA** outputs public key $PK = (pp, H_2, e(g,g)^\alpha, g^\beta)$ and the master key $MSK = (g^\alpha, \{h_x\}_{x \in A})$.

- **KeyGen**$_{DU}(PK, MSK, S)$. **TA** issues a set of attributes for each **DU** and randomly selects an element $v \in \mathbb{Z}_p$, and computes $K_1 = g^\alpha \cdot g^{v\beta}, K_2 = g^v, K_x = h_x^v (x \in S)$. **TA** issues $SK_u = (K_1, K_2, \{K_x\}_{x \in S})$ to **DU**.

- **KeyGen**$_{DM}(PK, \mathcal{O})$. **DM** randomly selects an element $c \in \mathbb{Z}_p$, and computes $PK_m = g^c$ and $SK_m = c$ as own public/private key pair. Given multiple owners $\mathcal{O} = \{O_1, \cdots, O_d\}$, **DM** outputs a $(d-1)$-dimensional polynomial $f(x) = a_0 + a_1 x + \cdots + a_{d-1} x^{(d-1)}$, where $a_i \in \mathbb{Z}_p (i \in [1, d-1])$, $a_0 = c$. Then, **DM** selects $d$ points $\{(x_1, y_1), \cdots, (x_d, y_d)\}$ according to $f(x)$ and sends $y_t (t \in [1, d])$ to each **DO**.

- **Enc**$(PK, (\mathbb{M}, \rho), F)$. Given a file $F$, **DM** encrypts the file $F$ into $C_F$ by using the symmetric secret key $K \in \mathbb{G}_T$. **DM** encrypts $K$ by a matrix $\mathbb{M}_{l \times q}$ associate with the access policy, the function $\rho(\tau)$ maps each row $\mathbb{M}_\tau (\tau \in [1, l])$ of the $\mathbb{M}_{l \times q}$ to an attribute, and **DM** randomly chooses a vector $\vec{v} = (s, y_2, \cdots, y_q)$ to compute $\lambda_\tau = \mathbb{M}_\tau \cdot \vec{v}$. **DM** randomly chooses $r_1, \cdots, r_l \in \mathbb{Z}_p$. The ciphertext of $K$ is:

$$CT_1 = \left( (\mathbb{M}, \rho), C_1 = K \cdot e(g,g)^{s\alpha}, C_2 = g^s, \left\{ C_\tau = g^{\beta \lambda_\tau} h_{\rho(\tau)}^{-r_\tau}, D_\tau = g^{r_\tau} \right\}_{\tau \in [1,l]} \right).$$

- **Sign**$(C_F, y_t)$. Each **DO** can generate the signature $\sigma_t = H_2(C_F)^{y_t}$ on the file after receiving $y_t$ from the **DM**.

- **Agg**($\{\sigma_t\}$). When **DM** receives the signatures of all **DOs**, **DM** computes the aggregated signature $\sigma = \prod_{t=1}^{d} \sigma_t^{L_t(0)} = (H_2(C_F))^c$, where $L_t(0) = \prod_{l=1, l\neq t}^{d} \frac{-x_l}{x_t - x_l}$. **DM** uploads the ciphertext $CT = \{C_F, CT_1, \sigma\}$ to **CSP**.
- **Verify**($PK, CT^*, PK_m$). When **DU** receives the ciphertext $CT^*$ from **CSP**, **DU** calculates $h^* = H_2(C_F^*)$. If $e(g, \sigma) = e(PK_m, h^*)$, **DU** executes the decryption operation. Otherwise, the algorithm will return $\bot$.
- **Dec**($CT^*, SK_u$). **DU** first needs to obtain the symmetric secret key $K$. If the attributes of **DU** satisfy the access policy embedded in the ciphertext, $K$ can be computed. Let $I \subset \{1,2,\cdots,l\}$ be defined as $I = \{\tau: \rho(\tau) \in S\}$, $\{\lambda_\tau\}$ is a reasonable share of $s$, and there exists a set of constants $\{\omega_\tau \in \mathbb{Z}_p\}_{\tau \in I}$ satisfying $\sum_{\tau \in I} \omega_\tau \lambda_\tau = s$. **DU** computes $A = \frac{e(c_2, k_1)}{\prod_{\tau \in I}(e(c_\tau, k_2)e(D_\tau, K_{\rho(\tau)}))^{\omega_\tau}}$, $K = C_1/A$. Finally, **DU** transforms the ciphertext $C_F^*$ into $F$ by the symmetric key $K$.
- **Update**($SK_m$). When $m$ new **DOs** join, $d^* = d + m$. **DM** recalls **KeyGen**$_{DM}$ algorithm to generate $PK_m^* = g^{c^*}, SK_m^* = c^*$. **DM** rechooses a $(d^* - 1)$ dimensional-polynomial $f^*(x) = a_0^* + a_1^* x + \cdots + a_{d^*-1}^* x^{(d^*-1)}$, where $a_{i'}^* \in \mathbb{Z}_p$, $a_0^* = c^*$, $(i' \in [1, d^* - 1])$. **DM** reselects $d^*$ points $\{(x_1^*, y_1^*), \cdots, (x_{d^*}^*, y_{d^*}^*)\}$ on $f^*(x)$ and sends $y_t^*(t \in [1, d^*])$ to each **DO**. Finally, **DM** computes the update key $UPK = c^*/c$ and uploads it to **CSP**. The new signature can be computed by $\sigma^* = \sigma^{UPK}$; When $n$ **DOs** leave, $d^* = d - n$, **DM** recalls the **KeyGen**$_{DM}$ algorithm to generate a new public/private key pair and reselects a $(d^* - 1)$ dimensional-polynomial $f^*(x)$. After that, the **DM** performs a process similar to the join operation.

### 5.3. Security analysis

#### 5.3.1. Correctness analysis

1) The correctness of **Verify** can be derived in the following way:
$e(g, \sigma) = e(g^c, H_2(C_F)) = e(GPK, h^*)$.
2) The correctness of **Dec** can be derived by:

$$A = \frac{e(C_2, K_1)}{\prod_{\tau \in I}(e(C_\tau, K_2)e(D_\tau, K_{\rho(\tau)}))^{\omega_\tau}}$$

$$= \frac{e(g^s, g^\alpha \cdot g^{v\beta})}{\prod_{\tau \in I} e(g^{\beta\lambda_\tau} h_{\rho(\tau)}^{-r_\tau}, g^v)^{\omega_\tau} e(g^{r_\tau}, h_{\rho(\tau)}^v)^{\omega_\tau}}$$

$$= e(g,g)^{s\alpha}$$

$$\frac{C_1}{A} = \frac{K \cdot e(g,g)^{s\alpha}}{e(g,g)^{s\alpha}} = K$$

#### 5.3.2. Security proof

***Theorem 1***: If the decisional q-parallel BDHE assumption holds, VDS-DM scheme is secure against selected plaintext attacks in the selection model.

***Proof 1***: Suppose there exists a probabilistic polynomial-time adversary that can compromise the security of the VDS-DM scheme with non-negligible probability, then there exists an algorithm $\mathcal{B}$ that can solve the decisional q-parallel BDHE assumption.

Initialization: $\mathcal{A}$ selects an access policy $P^*$, $\mathbb{M}^*$ is a matrix of $l^* \times q^*$ associated with the access policy, $\mathcal{A}$ sends $(\mathbb{M}^*, \rho^*)$ to $\mathcal{B}$.

***Setup***: $\mathcal{B}$ selects an element $\alpha'$ and sets $e(g,g)^\alpha = e(g^\beta, g^{\beta^q})e(g,g)^{\alpha'}$, here sets $\alpha = \alpha' + \beta^{q+1}$. $\mathcal{B}$ chooses $v_x \in \mathbb{Z}_p$ at random for each attribute $x$. Let $\tau$ denotes the row number of the matrix $\mathbb{M}^*$, $X$ denotes a set with index $\tau$, and $\rho^*(\tau) = x$. Then $\mathcal{B}$ computes $h_x = g^{v_x} \prod_{\tau \in X} g^{\beta \mathbb{M}^*_{\tau,1}/b_\tau} g^{\beta^2 \mathbb{M}^*_{\tau,2}/b_\tau} \cdots g^{\beta^{q^*} \mathbb{M}^*_{\tau,q^*}/b_\tau}$.

***Phase 1***: Suppose $\mathcal{B}$ receives a key query request for a set of attributes $S$, where $S$ does not satisfy the access policy $P^*$. $\mathcal{B}$ selects a vector $\vec{v}^* = \{v_1^*, v_2^*, \cdots, v_{q^*}^*\} \in \mathbb{Z}_p^*$, where $v_1^* = -1$. For every $\tau \in X, \rho^*(\tau) \in S$, there is $\mathbb{M}^*_\tau \vec{v}^* = 0$. $\mathcal{B}$ randomly chooses $r \in \mathbb{Z}_p$ and computes:

$$v = r + v_1^* \beta^q + v_2^* \beta^{q-1} + \cdots + v_{q^*}^* \beta^{q-q^*+1};$$

$$K_2 = g^r \prod_{\tau=1}^{q^*} \left(g^{\beta^{q-\tau+1}}\right)^{v_\tau^*} = g^v;$$

$$K_1 = g^{\alpha'} g^{\beta r} \prod_{\tau=2}^{q^*} \left(g^{\beta^{q-\tau+2}}\right)^{v_\tau^*} = g^\alpha g^{v\beta}.$$

For each attribute $x \in S$, if there is no $\tau$ such that $\rho^*(\tau) = x$, $\mathcal{B}$ sets $K_x = K_2^{v_x}$; Otherwise, $\mathcal{B}$ sets

$$K_x = K_2^{v_x} \prod_{\tau \in X} \prod_{j=1}^{q^*} \left(g^{r\frac{\beta^j}{b_\tau}} \prod_{k=1, k \neq j}^{q^*} \left(g^{\beta^{q+1+j-k}/b_\tau}\right)^{v_k^*}\right)^{\mathbb{M}^*_{\tau,j}} = h_x^v.$$

***Challenge***: $\mathcal{A}$ submits two equal-length messages $M_0, M_1$ to $\mathcal{B}$, $\mathcal{B}$ selects a random bit $b \in \{0,1\}$ and computes $C_1^* = K_b Te(g^s, g^{\alpha'})$, $C_2^* = g^s$. $\mathcal{B}$ randomly chooses $y_2', \cdots, y_{q^*}' \in \mathbb{Z}_p$ and sets $\vec{v}' = (s, s\beta + y_2', s\beta^2 + y_3', \cdots, s\beta^{q^*-1} + y_{q^*}')$. Let $Y$ denotes a set of $\tau$, where $\tau \in [1, l]$ and satisfies $\rho^*(\tau) = \rho^*(k)(k \neq \tau)$. $\mathcal{B}$ randomly chooses $r_1', \cdots, r_l' \in \mathbb{Z}_p$ and computes:

$$D_\tau^* = g^{-r_\tau' - sb_\tau};$$

$$C_\tau^* = h_{\rho^*(\tau)}^{r_\tau'} (g^{sb_\tau})^{-v_{\rho^*(\tau)}} \prod_{j=2}^{q^*} (g^\beta)^{y_j' \mathbb{M}^*_{\tau,j}} \prod_{k \in Y} \prod_{j=1}^{q^*} (g^{\beta^j s \frac{b_\tau}{b_k}})^{\mathbb{M}^*_{\tau,j}}.$$

***Guess***: $\mathcal{A}$ outputs a guess $b' \in \{0,1\}$.

If $b' = b$, $\mathcal{B}$ can guess that $T = e(g,g)^{\beta^{q+1}s}$, we have $\Pr[\mathcal{B}(\vec{y}, T = e(g,g)^{\beta^{q+1}s}) = 0] = \frac{1}{2} + Adv_\mathcal{A}^{CPA}$. Otherwise, $T$ is a random element in $\mathbb{G}_T$, we have $\Pr[\mathcal{B}(\vec{y}, T = R) = 0] = \frac{1}{2}$. The advantage that $\mathcal{B}$ can solve the decisional q-parallel BDHE assumption is negligible, then the VDS-DM scheme is selectively secure.

## 6. PERFORMANCE ANALYSIS

In this section, we evaluate the VDS-DM scheme from both theoretical and experimental aspects.

### 6.1. Theoretical analysis

We analyze the storage and calculation cost of the VDS-DM scheme theoretically. In storage cost, we set $|\mathbb{G}_T|, |\mathbb{G}|$ and $|\mathbb{Z}_p|$ to denote the length of element in $\mathbb{G}_T$, $\mathbb{G}$ and $\mathbb{Z}_p$. In calculation cost, we set the bilinear pairing operation $P$, the exponentiation operation $E(E_T)$ in group $\mathbb{G}$ ($\mathbb{G}_T$).

We ignore the hash operation which is more efficient compared to the above operations. We set $d$ to denote the number of **DOs**, $n_s$ to denote the number of attributes in the system, $n_a$ to denote the number of attributes in the access structure, $n_u$ to denote the number of **DU**'s attributes, and "-" to denote not applicable. Table 1 shows the calculation and storage cost of our scheme.

As shown in Table 1, in **Setup** algorithm, **TA** publishes a value for each system attribute. Therefore, the storage cost of this algorithm is affected by the number of system attributes. **TA** issues a set of attributes for **DU** and sends a secret key for **DU** according to attributes. Thus, the storage cost of **KeyGen$_{DU}$** algorithm is related to the number of **DU**'s attributes. The **KeyGen$_{DM}$** algorithm stores only two values fixedly, so its storage cost is constant. The ciphertext is related to the access policy, so the storage cost of **Enc** algorithm is affected by the number of attributes in the access policy. In the multi-owner setting, the storage cost of **Sign** algorithm is related to the number of owners. The **Agg** algorithm outputs an aggregated signature belonging to $\mathbb{G}$, so the storage cost is $|\mathbb{G}|$. The signature verification process does not store information, but only to perform verify operation, so the storage cost of **Verify** algorithm is not considered. **DU** needs to store the results of three linear pairing operations when executing **Dec** algorithm. Therefore, the storage cost is $3|\mathbb{G}_T|$.

Table 1. Storage and calculation cost

|  | Storage cost | Calculation cost |
| --- | --- | --- |
| **Setup** | $(3 + n_s)|\mathbb{G}| + |\mathbb{G}_T|$ | $P + 2E$ |
| **KeyGen$_{DU}$** | $(2 + n_u)|\mathbb{G}|$ | $(2 + n_u)E$ |
| **KeyGen$_{DM}$** | $|\mathbb{G}| + |\mathbb{Z}_p|$ | $E$ |
| **Enc** | $|\mathbb{G}_T| + (2n_a + 1)|\mathbb{G}|$ | $P + (2n_a + 1)E$ |
| **Sign** | $d|\mathbb{G}|$ | $E$ |
| **Agg** | $|\mathbb{G}|$ | $E$ |
| **Verify** | − | $2P$ |
| **Dec** | $3|\mathbb{G}_T|$ | $(2P + E_T)n_a + P$ |

In the calculation cost, the cost of **Setup** algorithm is not affected by the number of system attributes. In the composition of **DU**'s secret key, **TA** performs corresponding operation on the value corresponding to each attribute of **DU**. Therefore, the calculation cost of **KeyGen$_{DU}$** algorithm is related to the number of attributes of **DU**. **DM** obtains the public/private key pair through an exponentiation operation $E$ under **KeyGen$_{DM}$** algorithm. In **Sign** algorithm, each **DO** only perform an exponentiation operation $E$ to complete the signature, which reduces the calculation burden of resource-limited **DOs**. The calculation cost of **Agg** algorithm is independent of the number of **DO**. In **Verify** algorithm, the results of two bilinear pairing operation $P$ need to be judged. Thus, the calculation cost is $2P$. The calculation cost of **Enc** and **Dec** are related to the number of attributes in the access policy.

### 6.2. Experimental simulation

We test the performance of the VDS-DM scheme through experiment. Our experiment is implemented on a 64-bit Windows 10 operating system with an 11th Gen Intel(R) Core (TM) i7-11700T @ 1.40GHz1.39GHz processor. The experiment uses Java language and JPBC-1.2.1, and uses a-type curves based on 160-bit elliptic curve group on a super singular curve $y^2 = x^3 + x$ over a 512-bit finite field. we set $|\mathbb{Z}_p| = 160$bit, $|\mathbb{G}| = |\mathbb{G}_T| = 1024$bit, $d \in [2,10]$, $n_s$, $n_a$, $n_u \in [10,100]$.

We set the unit of storage cost to KB and the unit of calculation cost to *ms*. As shown in Figure 3, the storage cost of **Setup** is influenced by the number of system's attributes. Figure 4 shows the storage cost of **KeyGen**$_{DU}$ algorithm increases linearly with the number of **DU**'s attributes. From Figure 5, and compared with other algorithms, the storage cost of **Enc** algorithm is larger and increases with the number of attributes in the access structure. As shown in Figure 6, the calculation cost of the **KeyGen**$_{DU}$ algorithm is proportional to the number of **DU**'s attributes. In **KeyGen**$_{DU}$ algorithm, it takes about 339*ms* at $n_u = 50$. Figure 7 and Figure 8 show the calculation cost of **Enc** and **Dec**, both of which are linearly related to the number of attributes in the access structure. Among them, the **Enc** algorithm is more time consuming. Through experiments we found that each **DO** takes about 7*ms* to generate a signature by executing **Sign** algorithm in our experiment setting. The experiment shows that our scheme is feasible and efficient in solving the problem of verifiable data sharing in dynamic multi-owner setting.

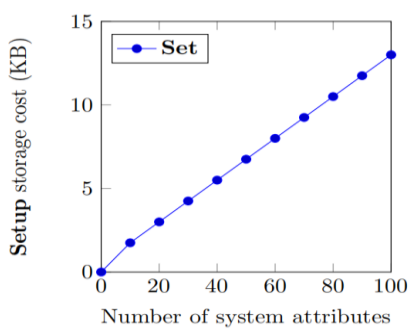

Figure 3. Storage cost in ***Setup***

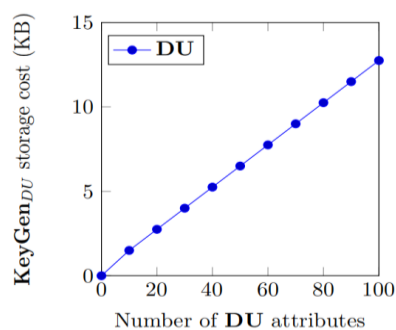

Figure 4. Storage cost in ***KeyGen***$_{DU}$

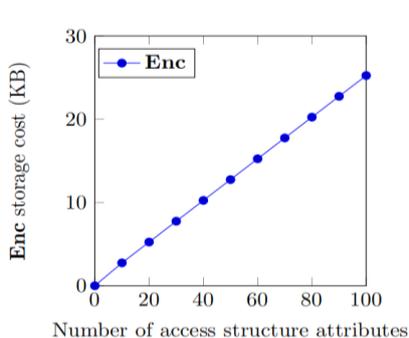

Figure 5. Storage cost in ***Enc***

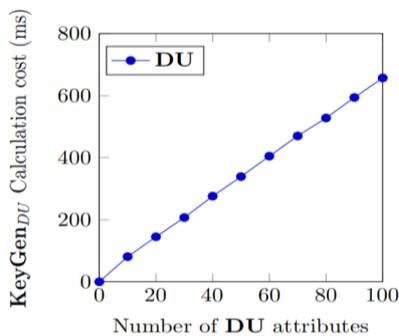

Figure 6. Calculation cost in ***KeyGen***$_{DU}$

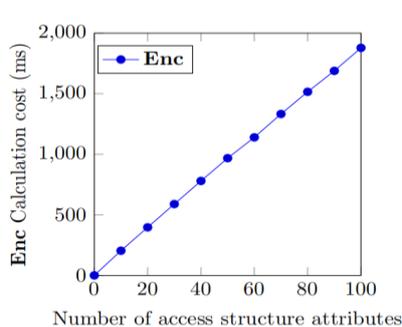

Figure 7. Calculation cost in ***Enc***

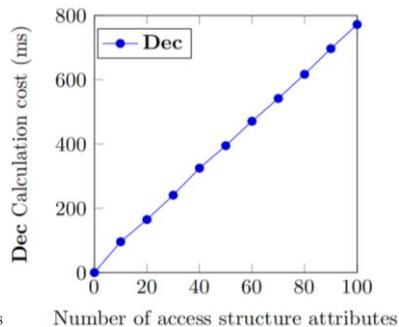

Figure 8. Calculation cost in ***Dec***

## 7. Conclusion

In this paper, we propose a verifiable data sharing scheme (called VDS-DM) that can support dynamic multi-owner scenarios, which can ensure the confidentiality of files and the privacy of user identities while achieving the verifiability of shared files. The proposed scheme can complete the update of the file permission signature without the assistance of a third party, addressing the update issue brought by the dynamic change of the file owners. This method reduces the communication overhead. Additionally, users can verify the integrity of shared files by themselves without resorting to a third party. We demonstrate the security of VDS-DM through a formal security game. Finally, we conduct sufficient simulation experiments, and the experimental results demonstrate the feasibility of VDS-DM.

### Acknowledgements

The authors would like to thank for the support from the Natural Science Foundation of Shandong Province under Grant No. ZR2022QF102.

### References

[1] A. Sunyaev, "Cloud computing," in *Internet computing*. Springer, 2020, pp. 195–236.
[2] C. Ge, W. Susilo, Z. Liu, J. Xia, P. Szalachowski, and L. Fang, "Secure keyword search and data sharing mechanism for cloud computing," *IEEE Transactions on Dependable and Secure Computing*, vol. 18, no. 6, pp. 2787–2800, 2020.
[3] J. Li, S. Wang, Y. Li, H. Wang, H. Wang, H. Wang, J. Chen, and Z. You, "An efficient attribute-based encryption scheme with policy update and file update in cloud computing," *IEEE Transactions on Industrial Informatics*, vol. 15, no. 12, pp. 6500–6509, 2019.
[4] S. Xu, J. Ning, Y. Li, Y. Zhang, G. Xu, X. Huang, and R. Deng, "Match in my way: Fine-grained bilateral access control for secure cloud-fog computing," *IEEE Transactions on Dependable and Secure Computing*, 2020.
[5] X. Lu, Z. Pan, and H. Xian, "An integrity verification scheme of cloud storage for internet-of-things mobile terminal devices," *Computers & Security*, vol. 92, p. 101686, 2020.
[6] X. Gao, J. Yu, Y. Chang, H. Wang, and J. Fan, "Checking only when it is necessary: Enabling integrity auditing based on the keyword with sensitive information privacy for encrypted cloud data," *IEEE Transactions on Dependable and Secure Computing*, 2021.
[7] D. Boneh, G. D. Crescenzo, R. Ostrovsky, and G. Persiano, "Public key encryption with keyword search," in *International conference on the theory and applications of cryptographic techniques*. Springer, 2004, pp. 506–522.
[8] R. Chen, Y. Mu, G. Yang, F. Guo, and X. Wang, "Dual-server publickey encryption with keyword search for secure cloud storage," *IEEE transactions on information forensics and security*, vol. 11, no. 4, pp. 789–798, 2015.
[9] J. Bethencourt, A. Sahai, and B. Waters, "Ciphertext-policy attributebased encryption," in *2007 IEEE symposium on security and privacy (SP'07)*. IEEE, 2007, pp. 321–334.
[10] Y. Miao, J. Ma, X. Liu, J. Weng, H. Li, and H. Li, "Lightweight finegrained search over encrypted data in fog computing," *IEEE Transactions on Services Computing*, vol. 12, no. 5, pp. 772–785, 2018.
[11] B. Waters, "Ciphertext-policy attribute-based encryption: An expressive, efficient, and provably secure realization," in *International workshop on public key cryptography*. Springer, 2011, pp. 53–70.
[12] Z. Zhang, J. Zhang, Y. Yuan, and Z. Li, "An expressive fully policyhidden cipher text policy attribute-based encryption scheme with credible verification based on blockchain," *IEEE Internet of Things Journal*, 2021.
[13] M. Xiao, H. Li, Q. Huang, S. Yu, and W. Susilo, "Attribute-based hierarchical access control with extendable policy," *IEEE Transactions on Information Forensics and Security*, 2022.
[14] Y. Miao, J. Ma, X. Liu, X. Li, Q. Jiang, and J. Zhang, "Attributebased keyword search over hierarchical data in cloud computing," *IEEE Transactions on Services Computing*, vol. 13, no. 6, pp. 985–998, 2017.

[15] Z. Ying, W. Jiang, X. Liu, S. Xu, and R. Deng, "Reliable policy updating under efficient policy hidden fine-grained access control framework for cloud data sharing," *IEEE Transactions on Services Computing*, 2021.

[16] D. Han, N. Pan, and K.-C. Li, "A traceable and revocable cipher text policy attribute-based encryption scheme based on privacy protection," *IEEE Transactions on Dependable and Secure Computing*, 2020.

[17] N. Chen, J. Li, Y. Zhang, and Y. Guo, "Efficient cp-abe scheme with shared decryption in cloud storage," *IEEE Transactions on Computers*, vol. 71, no. 1, pp. 175–184, 2020.

[18] J. Lai, R. H. Deng, C. Guan, and J. Weng, "Attribute-based encryption with verifiable outsourced decryption," *IEEE Transactions on information forensics and security*, vol. 8, no. 8, pp. 1343–1354, 2013.

[19] B. Qin, R. H. Deng, S. Liu, and S. Ma, "Attribute-based encryption with efficient verifiable outsourced decryption," *IEEE Transactions on Information Forensics and Security*, vol. 10, no. 7, pp. 1384–1393, 2015.

[20] Y. Miao, R. H. Deng, K.-K. R. Choo, X. Liu, J. Ning, and H. Li, "Optimized verifiable fine-grained keyword search in dynamic multi-owner settings," *IEEE Transactions on Dependable and Secure Computing*, vol. 18, no. 4, pp. 1804–1820, 2019.

[21] Y. Zhang, T. Zhu, R. Guo, S. Xu, H. Cui, and J. Cao, "Multi-keyword searchable and verifiable attribute-based encryption over cloud data," *IEEE Transactions on Cloud Computing*, 2021.

[22] A. Sahai and B. Waters, "Fuzzy identity-based encryption," in *Annual international conference on the theory and applications of cryptographic techniques*. Springer, 2005, pp. 457–473.

[23] A. Shamir, "Identity-based cryptosystems and signature schemes," in *Workshop on the theory and application of cryptographic techniques*. Springer, 1984, pp. 47–53.

[24] V. Goyal, O. Pandey, A. Sahai, and B. Waters, "Attribute-based encryption for fine-grained access control of encrypted data," in *Proceedings of the 13th ACM conference on Computer and communications security*, 2006, pp. 89–98.

[25] Y. Miao, J. Ma, X. Liu, J. Zhang, and Z. Liu, "Vkse-mo: verifiable keyword search over encrypted data in multi-owner settings," *Science China Information Sciences*, vol. 60, no. 12, pp. 1–15, 2017.

[26] Y. Miao, X. Liu, K.-K. R. Choo, R. H. Deng, J. Li, H. Li, and J. Ma, "Privacy-preserving attribute-based keyword search in shared multi-owner setting," *IEEE Transactions on Dependable and Secure Computing*, vol. 18, no. 3, pp. 1080–1094, 2019.

[27] W. Sun, S. Yu, W. Lou, Y. T. Hou, and H. Li, "Protecting your right: Verifiable attribute-based keyword search with fine-grained ownerenforced search authorization in the cloud," *IEEE Transactions on Parallel and Distributed Systems*, vol. 27, no. 4, pp. 1187–1198, 2014.

## AUTHORS

**Jing Zhao** received the B.S. degree from the College of Electronic Information, Shandong Modern University, Jinan, China, in 2021. She is currently a graduate student in the College of Computer Science and Technology, Qingdao University, Qingdao, China. Her field of study is Attribute-Based Encryption.

**Qianqian Su** received the B.S. and M.S. degrees from the College of Computer Science and Technology, Qingdao University, Qingdao, China, in 2015 and 2018, respectively. She received a Ph.D. degree from the State Key Laboratory of Information Security, Institute of Information Engineering, Chinese Academy of Sciences, Beijing, China, in 2021. She is currently working in the School of Computer Science and Technology, Qingdao University, Qingdao, China. Her research interests include cloud computing security, blockchain, and data security.